\documentclass[
  ]{article}

\usepackage{arxiv}

\usepackage{siunitx}               
\usepackage{tikz, adjustbox, tikz-qtree}
\usepackage[protrusion=false]{microtype}

\usepackage[utf8]{inputenc} 
\usepackage[T1]{fontenc}    
\usepackage{hyperref}       
\usepackage{url}            
\usepackage{booktabs}       
\usepackage{longtable}      
\usepackage{multirow}       
\usepackage{listings}       
\usepackage{amsfonts}       
\usepackage{amsmath}        

\usepackage{nicefrac}       
\usepackage{microtype}      
\usepackage{lipsum}		      
\usepackage{bm}             
\usepackage{graphicx}       
\usepackage{adjustbox}      
\usepackage{orcidlink}      
\usepackage{xstring}        
\usepackage{datetime2}      
\usepackage{xparse}         
\usepackage{calc}           
\usepackage{svg-extract}    

\newcommand{\passthrough}[1]{#1}

\providecommand{\tightlist}{%
  \setlength{\itemsep}{0pt}\setlength{\parskip}{0pt}}
\NewDocumentCommand\citeproctext{}{}
\NewDocumentCommand\citeproc{mm}{%
  \begingroup\def\citeproctext{#2}\cite{#1}\endgroup}
\makeatletter
 \let\@cite@ofmt\@firstofone
 \def\@biblabel#1{}
 \def\@cite#1#2{{#1\if@tempswa , #2\fi}}
\makeatother

\makeatletter
\def\maxwidth{\ifdim\Gin@nat@width>\linewidth\linewidth\else\Gin@nat@width\fi}
\def\maxheight{\ifdim\Gin@nat@height>\textheight\textheight\else\Gin@nat@height\fi}
\makeatother
\setkeys{Gin}{width=\maxwidth,height=\maxheight,keepaspectratio}

\newcommand{\formatteddate}{2025-11-18}
\newlength{\cslhangindent}
\newlength{\csllabelwidth}
\newenvironment{CSLReferences}[2] 
  {\begin{list}{}{%
   \setlength{\itemindent}{0pt}
   \setlength{\leftmargin}{0pt}
   \setlength{\parsep}{0pt}
   \ifodd #1
    \setlength{\leftmargin}{\cslhangindent}
    \setlength{\itemindent}{-1\cslhangindent}
   \fi
   \setlength{\itemsep}{#2\baselineskip}}}
  {\end{list}}

\newcommand{\CSLLeftMargin}[1]{\parbox[t]{\csllabelwidth}{\strut#1\strut}}
\newcommand{\CSLRightInline}[1]{\parbox[t]{\linewidth - \csllabelwidth}{\strut#1\strut}}

\usepackage{algorithm2e}
\usepackage{tikz}

\usepackage[capitalise]{cleveref}  

    \title{HBAT 2: A Python Package to Analyse Hydrogen Bonds and Other
Non-covalent Interactions in Macromolecular Structures}

\date{\formatteddate}

\author{
  Abhishek Tiwari~\orcidlink{0000-0003-2222-2395}\\
      Independent Researcher\\
    \texttt{abhishek@abhishek-tiwari.com}
  }

\begin{document}

\renewcommand{\arxivdoi}{\href{https://doi.org/10.5281/zenodo.17645321}{zenodo.17645321}}
\renewcommand{\arxivauthor}{Abhishek Tiwari}

\maketitle

\setcounter{section}{0}

\setcounter{secnumdepth}{5}
\renewcommand{\thesection}{\arabic{section}}
\renewcommand{\thesubsection}{\thesection.\arabic{subsection}}
\renewcommand{\thesubsubsection}{\thesubsection.\arabic{subsubsection}}
\renewcommand{\theparagraph}{\thesubsubsection.\arabic{paragraph}}
\renewcommand{\thesubparagraph}{\theparagraph.\arabic{subparagraph}}

\section{Abstract}\label{abstract}

Hydrogen bonds and other non-covalent interactions play a crucial role
in maintaining the structural integrity and functionality of biological
macromolecules such as proteins and nucleic acids. Accurate
identification and analysis of these interactions are essential for
understanding molecular recognition, protein folding, and drug design.
HBAT\footnote{\url{https://hbat.abhishek-tiwari.com}} (Hydrogen Bond
Analysis Tool) is software for analysing hydrogen bonds and other weak
interactions in macromolecular structures. This paper presents HBAT 2,
an updated Python reimplementation of the original HBAT tool published
in 2007. HBAT 2 is a Python package for automated analysis of hydrogen
bonds and other non-covalent interactions in macromolecular structures
available in Protein Data Bank (PDB) file format. The software
identifies and analyses traditional hydrogen bonds, weak hydrogen bonds,
halogen bonds, X-H\(\cdots\)\(\pi\), \(\pi\)-\(\pi\) stacking, and
n\(\rightarrow\)\(\pi\)* interactions using geometric criteria. It also
detects cooperativity and anticooperativity chains and renders them as
2D visualisations. The latest version offers improved cross-platform
\passthrough{\lstinline!tkinter!}-based graphical user interface (GUI),
a web-based interface\footnote{\url{https://hbat-web.abhishek-tiwari.com}},
a simple command-line interface (CLI), and a developer-friendly API,
making it accessible to users with different computational backgrounds.

\section{Keywords}\label{keywords}

Python, structural biology, hydrogen bonds, molecular interactions,
protein structures, bioinformatics, non-covalent interactions, PDB
analysis

\section{Introduction}\label{introduction}

Hydrogen bonds and other non-covalent interactions are fundamental to
protein structure, stability, and function. With over 200,000 structures
in the Protein Data Bank \citeproc{ref-berman2000protein}{{[}1{]}},
there is an increasing need for automated tools to analyse these
interactions systematically.

The landscape of hydrogen bond analysis tools is diverse but fragmented.
Classic tools like HBPLUS \citeproc{ref-mcdonald1994satisfying}{{[}2{]}}
and HBexplore \citeproc{ref-lindauer1996hbexplore}{{[}3{]}} pioneered
automated H-bond detection but lack modern interfaces and support for a
diverse range of interactions. More recent tools serve specialized
niches: PLIP \citeproc{ref-salentin_plip_2015}{{[}4{]}} and Arpeggio
\citeproc{ref-jubb_arpeggio_2017}{{[}5{]}} excel at protein-ligand
interactions but are web-based without standalone GUI options; HBonanza
\citeproc{ref-durrant_hbonanza_2011}{{[}6{]}}, HBCalculator
\citeproc{ref-wang_hbcalculator_2024}{{[}7{]}}, and BRIDGE2
\citeproc{ref-siemers_interactive_2021}{{[}8{]}} focus on molecular
dynamics trajectories rather than static structures; MDAnalysis
\citeproc{ref-noauthor_431_nodate}{{[}9{]}}, GROMACS
\citeproc{ref-noauthor_gmx_nodate}{{[}10{]}}, and AMBER
\citeproc{ref-noauthor_hbond_2020}{{[}11{]}} provide H-bond analysis
within larger MD suites. Tools like VMD
\citeproc{ref-noauthor_vmd_nodate}{{[}12{]}} and ChimeraX
\citeproc{ref-noauthor_tool_nodate}{{[}13{]}} offer interactive hydrogen
bond visualisation but limited statistical analysis. ProteinTools
\citeproc{ref-ferruz_proteintools_2021}{{[}14{]}} provides web-based
network analysis but lacks cross-platform desktop capabilities.

Despite this ecosystem, there remains a gap for a comprehensive and
reliable cross-platform desktop tool that: (1) analyses diverse
interaction types beyond canonical hydrogen bonds, (2) provides both
graphical and command-line interfaces for different user workflows, (3)
identifies potential cooperativity, (4) integrates seamlessly with the
existing scientific ecosystem, and (5) supports flexible parameter
customization with domain-specific presets.

The original HBAT \citeproc{ref-tiwari2007hbat}{{[}15{]}} was developed
in Perl/Tk with a Windows-only GUI, limiting its adoption in modern
computational environments. HBAT 2 addresses these limitations by
providing a modern, Python implementation that integrates seamlessly
with contemporary structural biology workflows. The software is
particularly valuable for researchers in structural biology,
computational chemistry, and drug design who need detailed analysis of
molecular interactions.

\begin{figure}
\centering
\includegraphics{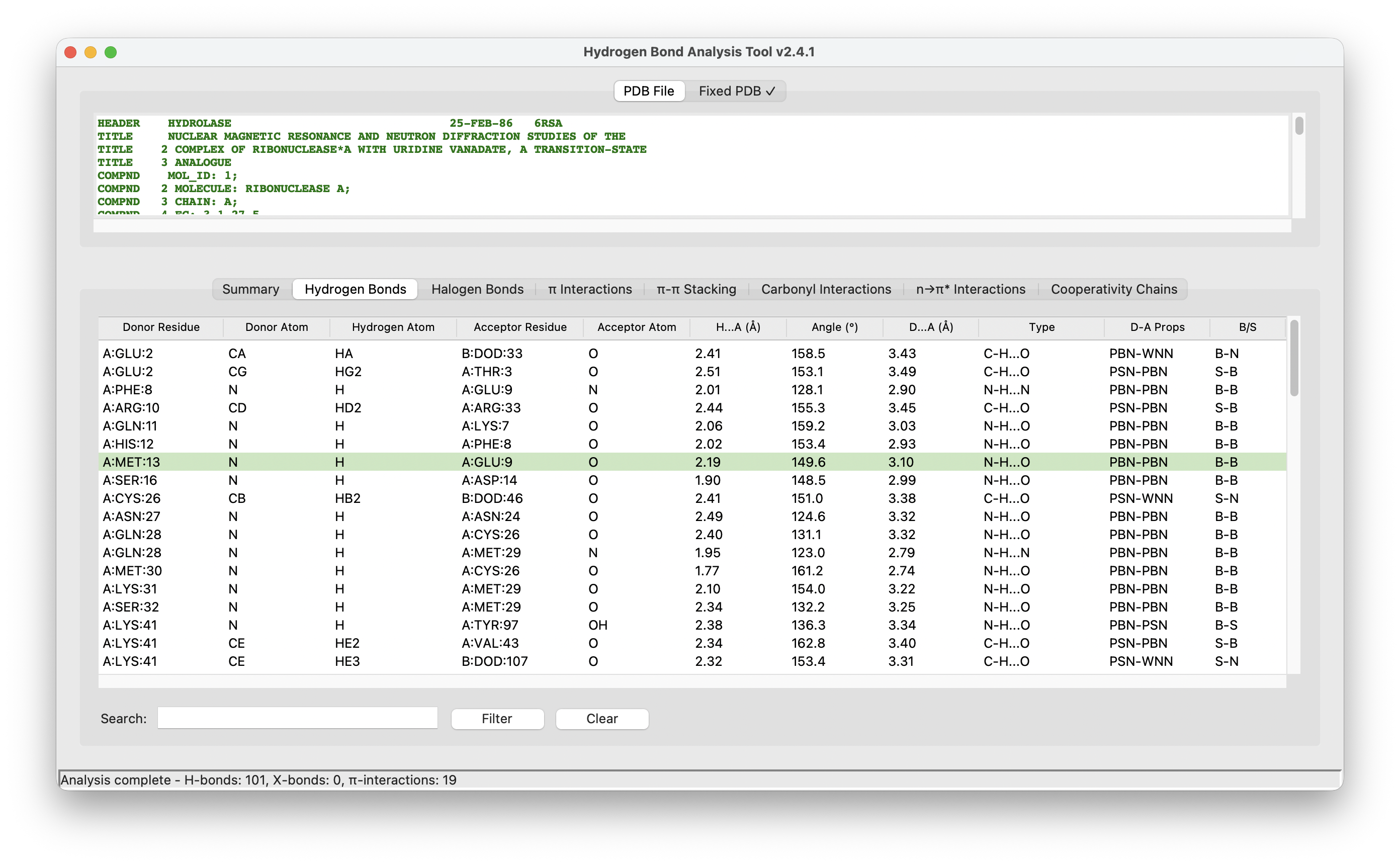}
\caption{The latest update to HBAT 2 uses tkinter to provide a
cross-platform graphical user interface (GUI)}
\end{figure}

\section{Methods}\label{methods}

\subsection{Implementation}\label{implementation}

HBAT 2 employs a modular architecture with separate components for PDB
parsing, geometric analysis, statistical computation, and visualisation.
The core analysis engine uses efficient nearest-neighbor searching with
configurable distance cutoffs, followed by geometric filtering based on
distance and angular criteria.

\begin{figure}
\centering
\includegraphics{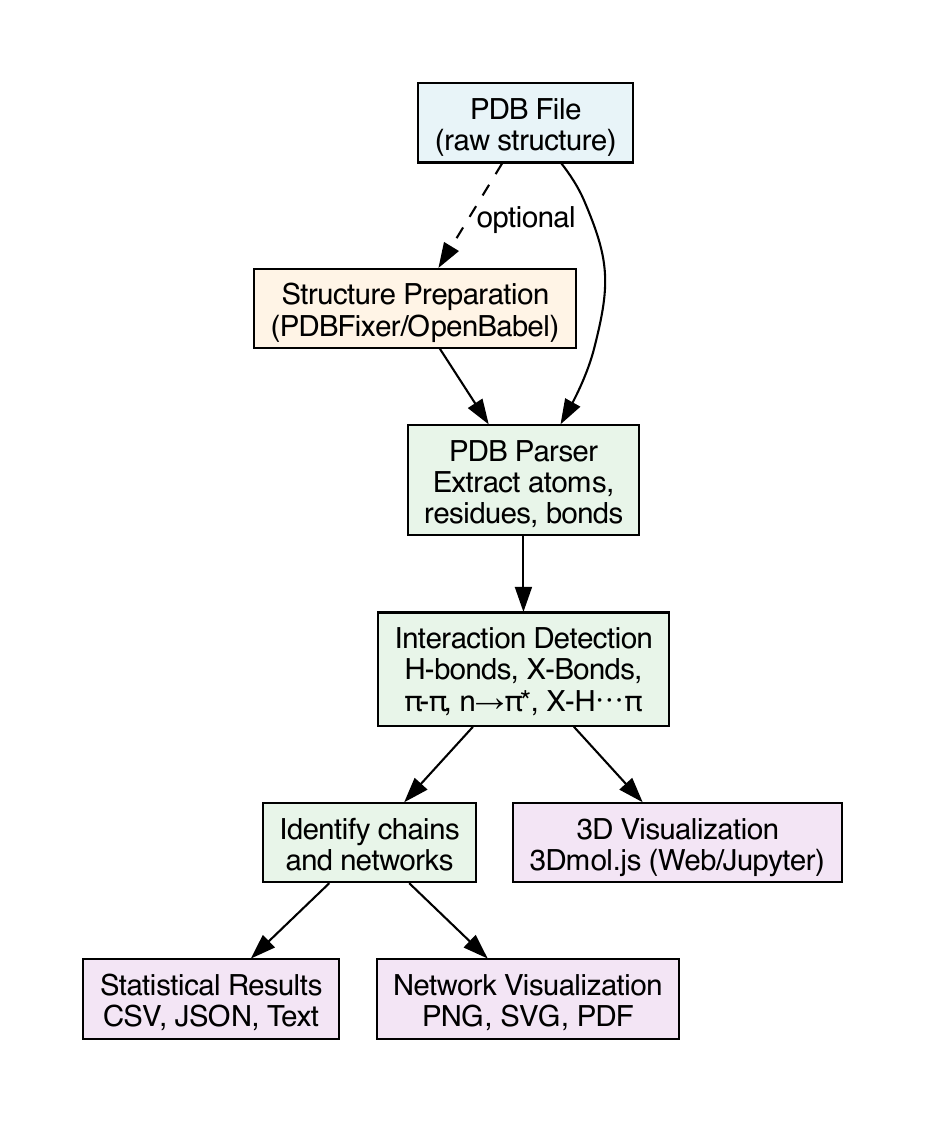}
\caption{HBAT 2 analysis workflow showing the complete pipeline from PDB input to multiple output formats}\label{fig-workflow}
\end{figure}

\subsubsection{Structure Preparation}\label{structure-preparation}

HBAT 2 uses PDBFixer
\citeproc{ref-noauthor_openmmpdbfixer_2025}{{[}16{]}},
\citeproc{ref-eastman_openmm_2013}{{[}17{]}} and OpenBabel
\citeproc{ref-oboyle_pybel_2008}{{[}18{]}} to automatically enhance
macromolecular structures by adding missing atoms, converting residues,
and cleaning up structural issues. These capabilities are particularly
valuable when working with crystal structures missing hydrogen atoms,
low-resolution structures with incomplete side chains, structures
containing non-standard amino acid residues, and structures with
unwanted ligands or contaminants.

\subsubsection{Interaction Detection}\label{interaction-detection}

The software implements the same fundamental geometric approach as the
original version \citeproc{ref-tiwari2007hbat}{{[}15{]}} but with
optimised algorithms and improved handling. HBAT 2 analyses a broad
spectrum of interactions including:

\begin{itemize}
\tightlist
\item
  Hydrogen bonds (O-H\(\cdots\)O, N-H\(\cdots\)O, N-H\(\cdots\)N,
  C-H\(\cdots\)O)
\item
  Halogen bonds (C-X\(\cdots\)Y where X=F,Cl,Br,I)
\item
  X-H\(\cdots\)\(\pi\) interactions with aromatic systems
\item
  \(\pi\)-\(\pi\) stacking
  \citeproc{ref-mcgaughey_pi-stacking_1998}{{[}19{]}},
  \citeproc{ref-vernon_pi-pi_nodate}{{[}20{]}}
\item
  Carbonyl-carbonyl n\(\rightarrow\)\(\pi\)* interactions
  \citeproc{ref-rahim_reciprocal_2017}{{[}21{]}},
  \citeproc{ref-newberry_n_2017}{{[}22{]}}
\item
  n\(\rightarrow\)\(\pi\)* interactions
  \citeproc{ref-choudhary_nature_2009}{{[}23{]}}
\end{itemize}

This comprehensive approach addresses the growing recognition of weak
interactions' importance in protein structure and stability
\citeproc{ref-desiraju_weak_2001}{{[}24{]}},
\citeproc{ref-cavallo_halogen_2016}{{[}25{]}},
\citeproc{ref-brandl_c-h-interactions_2001}{{[}26{]}}.

\subsubsection{Cooperativity Analysis}\label{cooperativity-analysis}

HBAT 2 offers two ways to visualise hydrogen bond networks: NetworkX
\citeproc{ref-hagberg2008networkx}{{[}27{]}}/Matplotlib
\citeproc{ref-hunter2007matplotlib}{{[}28{]}} and GraphViz
\citeproc{ref-graphviz2024}{{[}29{]}}. Unlike tools that provide only
basic visualisation (VMD, ChimeraX) or focus solely on MD trajectory
dynamics (BRIDGE2, HBonanza), HBAT 2 emphasises cooperativity chains and
network topology in static structures with customizable layouts and
high-resolution export (PNG, SVG, PDF). Cooperativity analysis and
visualisation support in HBAT 2 builds on top of original HBAT
\citeproc{ref-tiwari2007hbat}{{[}15{]}} and HBNG
\citeproc{ref-tiwari_hbng_2007}{{[}30{]}}.

\subsubsection{3D Visualization}\label{d-visualization}

HBAT 2 integrates 3Dmol.js \citeproc{ref-rego_3dmol_2014}{{[}31{]}} for
interactive 3D visualization of molecular structures and detected
interactions. This JavaScript-based viewer is available both on the web
server \footnote{\url{https://hbat-web.abhishek-tiwari.com}} and as a
widget in Jupyter notebooks, enabling researchers to interactively
explore structures without requiring separate visualization software.
The 3D viewer highlights detected interactions with customizable color
schemes and allows users to rotate, zoom, and inspect specific
interaction geometries. In Jupyter notebooks, the visualization widget
can be embedded directly alongside analysis code, facilitating
reproducible computational workflows and interactive data exploration.
This integration bridges the gap between automated analysis and manual
structural inspection, allowing users to validate detected interactions
visually and gain intuitive understanding of complex interaction
networks. Shortcuts to launch example Jupyter notebooks with Google
Colab - a hosted Jupyter Notebook service that requires no setup to use
and provides free access to computing resources - are also
provided\footnote{\url{https://github.com/abhishektiwari/hbat/tree/main/notebooks}}.

\begin{figure}
\centering
\includegraphics{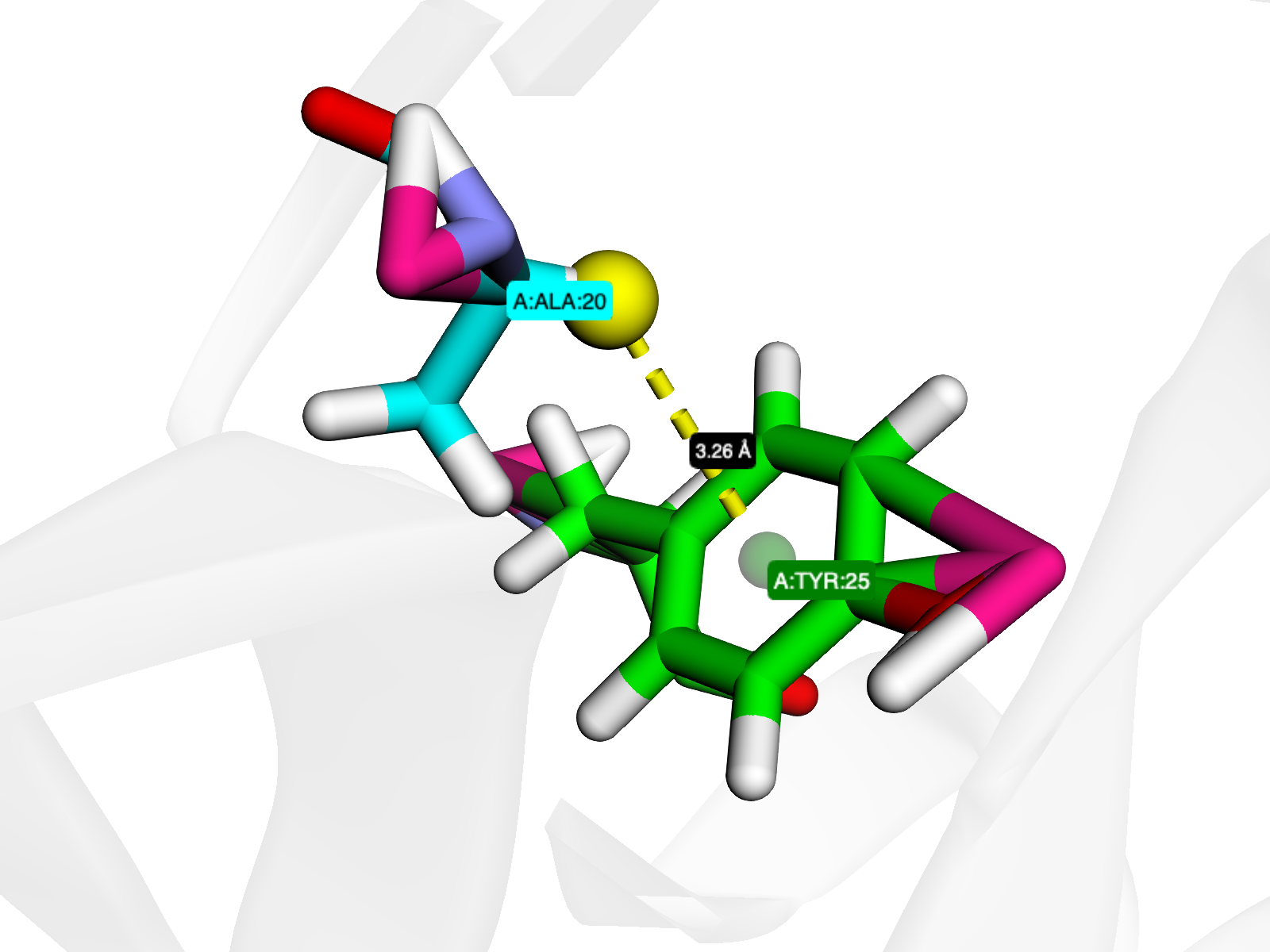}
\caption{C-H\(\cdots\)\(\pi\) interaction between A:ALA:20 → A:TYR:25 of
PDB Structure 6RSA}
\end{figure}

\subsubsection{Parameter Presets}\label{parameter-presets}

Built-in parameter sets optimised for different experimental conditions
(high-resolution X-ray, NMR, membrane proteins, drug design) address a
common challenge in hydrogen bond analysis. While other tools require
manual parameter specification, HBAT's preset system makes it accessible
to experimental structural biologists while maintaining flexibility for
computational experts.

\subsubsection{Output Formats}\label{output-formats}

Multiple export formats (text, CSV, JSON) enable integration with
downstream analysis pipelines and statistical software. Combined with
optional PDBFixer and OpenBabel integration for automated hydrogen
addition, HBAT 2 provides a complete workflow from raw PDB files to
publication-ready analyses.

\subsection{Operation}\label{operation}

HBAT 2 requires Python 3.8 or higher and runs on Windows, macOS, and
Linux. The software can be installed via pip
(\passthrough{\lstinline!pip install hbat!}) or conda
(\passthrough{\lstinline!conda install -c hbat hbat!}). Optional
dependencies include GraphViz for advanced network visualization and
PDBFixer/OpenBabel for structure preparation.

A web-based interface is available at
\url{https://hbat-web.abhishek-tiwari.com} providing immediate access
without local installation. The web server supports all core analysis
features including structure preparation, interaction detection, and
cooperativity analysis with interactive visualization. Users can upload
PDB files up to 1MB and download results in multiple formats (CSV, JSON,
and fixed PDB files).

\begin{figure}
\centering
\includegraphics{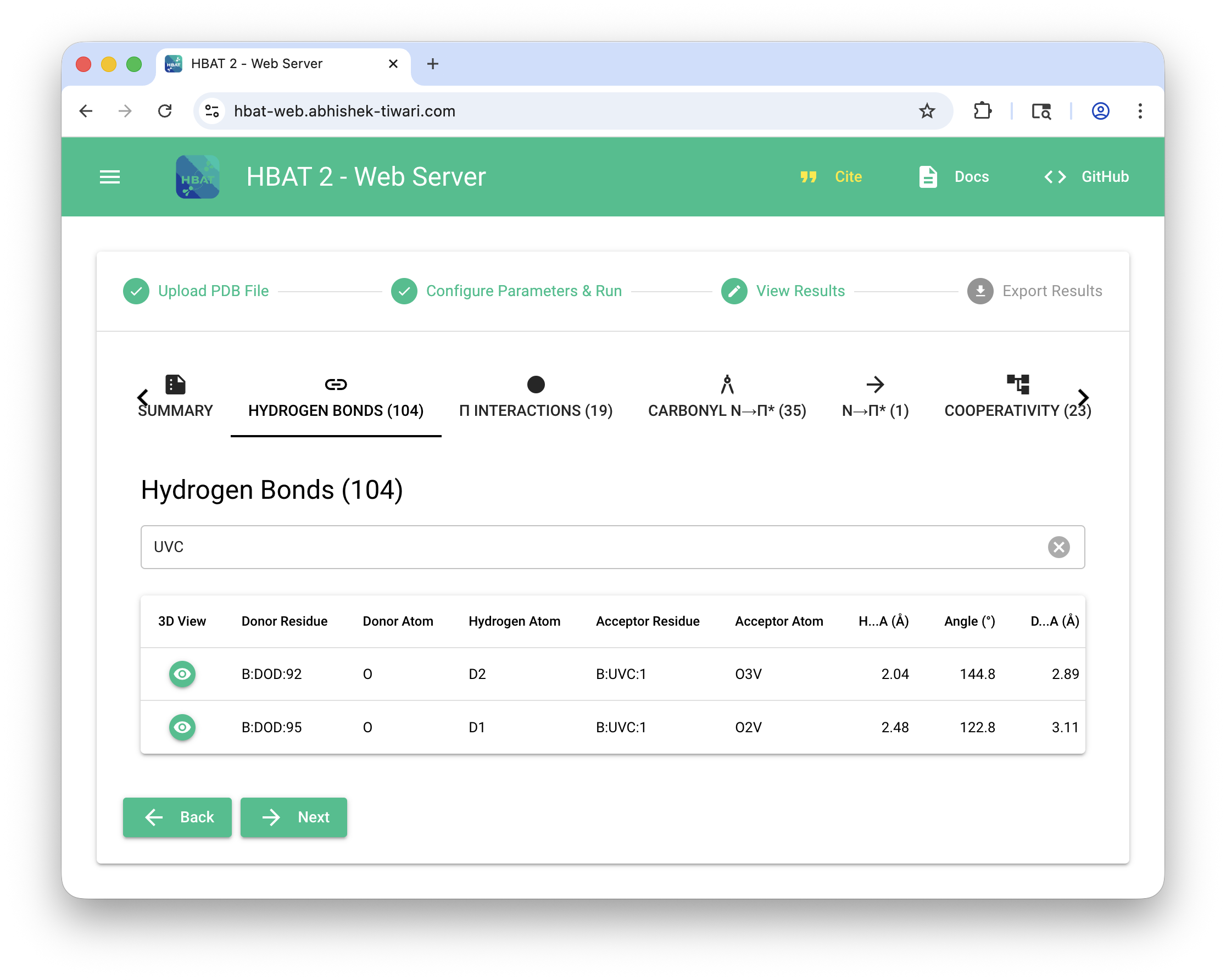}
\caption{HBAT 2 is also available as a web server and supports analysis
of PDB files up to 1MB.}
\end{figure}

\section{Use Cases}\label{use-cases}

HBAT 2 extends the impact of the original tool by providing modern
capabilities across diverse research areas. Since its initial
publication, HBAT has been employed in numerous studies spanning
structural biology, drug discovery, and clinical genomics.

\subsection{Structure-based drug
design}\label{structure-based-drug-design}

Analysis of protein-ligand interactions including hydrogen bonds and
halogen bonds crucial for binding affinity and rational inhibitor
design.

\begin{itemize}
\tightlist
\item
  Díaz-Cervantes et al.~characterized protein-ligand interactions for
  anticancer compounds targeting ABL kinase
  \citeproc{ref-diaz-cervantes_molecular_2020}{{[}32{]}}
\item
  Ravindran et al.~analyzed cardiovascular disease receptors with
  Withaferin A metabolite interactions
  \citeproc{ref-ravindran_interaction_2015}{{[}33{]}}
\item
  Zhou et al.~predicted EGFR drug resistance based on hydrogen bond
  patterns \citeproc{ref-zhou_prediction_2013}{{[}34{]}}
\item
  Russo Krauss et al.~elucidated thrombin-aptamer interactions crucial
  for anticoagulant design
  \citeproc{ref-russo_krauss_thrombinaptamer_2011}{{[}35{]}}
\end{itemize}

\subsection{Protein engineering}\label{protein-engineering}

Identification of stabilizing interactions for rational mutagenesis to
improve enzyme thermostability, activity, and specificity.

\begin{itemize}
\tightlist
\item
  Wang et al.~identified weak C-H\(\cdots\)\(\pi\) and C-H\(\cdots\)O
  interactions in thermostable \(\alpha\)-amylase variants, revealing
  that mutations in central \(\beta\)-strands can enhance enzyme
  performance \citeproc{ref-wang_simultaneously_2020}{{[}36{]}}
\item
  Fournier et al.~characterized hydrogen bonding networks essential for
  halide specificity in vanadium iodoperoxidases
  \citeproc{ref-fournier_vanadium_2014}{{[}37{]}}
\end{itemize}

\subsection{Molecular dynamics
analysis}\label{molecular-dynamics-analysis}

Statistical analysis of interaction patterns across molecular dynamics
trajectories and time-resolved structural ensembles.

\begin{itemize}
\tightlist
\item
  Huang, Massa, and Karle (Nobel laureate Jerome Karle) characterized
  hydrogen bonds in vesicular stomatitis virus nucleoprotein, enabling
  quantum mechanical energy calculations critical for understanding
  viral RNA encapsulation \citeproc{ref-huang_kernel_2009}{{[}38{]}}
\item
  Jayaprakash et al.~analyzed glycoprotein-lectin interactions for ER
  quality control mechanisms
  \citeproc{ref-jayaprakash_atomic_2025}{{[}39{]}}
\item
  Barzegar and Tohidifar investigated Hoogsteen H-bond stabilization in
  G-quadruplex DNA nanomotors
  \citeproc{ref-barzegar_stabilization_2025}{{[}40{]}}
\end{itemize}

\subsection{Crystallographic studies}\label{crystallographic-studies}

Quality assessment and validation of refined structures through detailed
hydrogen bond network analysis.

\begin{itemize}
\tightlist
\item
  Daubner et al.~revealed SRSF2's novel recognition mechanism for
  guanine and cytosine through syn-anti conformational flexibility in
  NMR structures \citeproc{ref-daubner_synanti_2012}{{[}41{]}}
\end{itemize}

\subsection{Comparative structural
analysis}\label{comparative-structural-analysis}

Systematic comparison of interaction networks across protein families to
understand structural effects of mutations and polymorphisms.

\begin{itemize}
\tightlist
\item
  Kavitha and Mohanapriya identified deleterious TOP2A mutations in
  ovarian cancer \citeproc{ref-kavitha_insights_2024}{{[}42{]}}
\item
  Khan and Ansari predicted highly deleterious E17K mutation in AKT1
  gene \citeproc{ref-khan_prediction_2017}{{[}43{]}}
\item
  Khan et al.~characterized functional SNPs in Axin 1 gene associated
  with colorectal cancer
  \citeproc{ref-khan_identification_2018}{{[}44{]}}
\item
  Abdulazeez et al.~identified rs61742690 (S783N) SNP as target for
  disrupting BCL11A-mediated fetal-to-adult globin switching
  \citeproc{ref-abdulazeez_rs61742690_2019}{{[}45{]}}
\item
  AbdulAzeez and Borgio analyzed deleterious nsSNPs in HBA1 gene
  associated with \(\alpha\)-thalassemia
  \citeproc{ref-abdulazeez_-silico_2016}{{[}46{]}}
\end{itemize}

\begin{figure}
\centering
\includegraphics{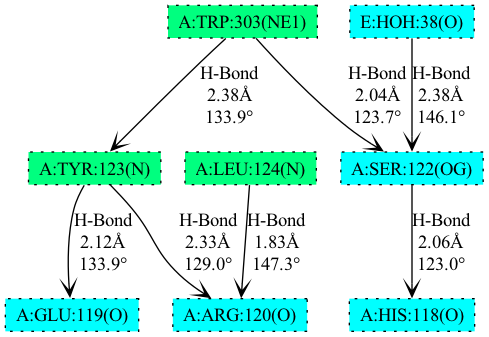}
\caption{An example visualisation of potential cooperativity chain
generated by HBAT 2 software for Protein Data Bank (PDB) entry 4X21}
\end{figure}

The software's preset configurations and flexible parameter system make
it accessible to both computational experts and experimental structural
biologists. Since its original publication, HBAT has been
\href{https://scholar.google.com/citations?view_op=view_citation&hl=en&user=Mb7eYKYAAAAJ&citation_for_view=Mb7eYKYAAAAJ:u-x6o8ySG0sC}{cited}
in numerous studies of protein structure and molecular recognition
\citeproc{ref-tiwari2007hbat}{{[}15{]}}.

\section{Discussion}\label{discussion}

HBAT 2 represents a significant modernization of the original HBAT tool,
addressing key limitations while preserving its core functionality. The
transition to Python and addition of cross-platform GUI support removes
barriers to adoption that limited the original Perl/Tk implementation.
The expanded repertoire of interaction types and cooperativity analysis
capabilities position HBAT 2 as a comprehensive tool for modern
structural biology research.

The modular architecture and well-documented API facilitate integration
into larger computational pipelines and enable customization for
specialized research needs. The preset parameter system, informed by
best practices from structural biology literature, lowers the barrier to
entry for experimental researchers while maintaining the flexibility
required by computational specialists.

\subsection{Limitations}\label{limitations}

Current limitations of HBAT 2 include: (1) analysis focuses on static
structures without accounting for conformational dynamics or solvent
effects, (2) interaction detection relies on geometric criteria alone
without energy calculations, which may miss energetically unfavorable
but geometrically valid interactions, (3) network visualization
performance may degrade with very large structures containing
\textgreater5,000 interactions, (4) preset parameters are optimized
primarily for protein structures and may require adjustment for other
biomolecules such as DNA/RNA complexes, and (5) cooperativity analysis
identifies potential chains based on topology but does not quantify
cooperative energetic effects.

Future development will focus on enhanced support for large biomolecular
complexes, integration with molecular dynamics analysis workflows, and
machine learning-based interaction prediction to complement geometric
criteria.

\section{Data Availability}\label{data-availability}

No new data were generated for this software tool article. Example PDB
files used for validation are available from the Protein Data Bank
(https://www.rcsb.org/).

\section{Software Availability}\label{software-availability}

\begin{itemize}
\tightlist
\item
  HBAT web server available on:
  \url{https://hbat-web.abhishek-tiwari.com/}
\item
  HBAT documentation available on:
  \url{https://hbat.abhishek-tiwari.com/}
\item
  Source code available from:
  \url{https://github.com/abhishektiwari/hbat}
\item
  PyPI Package available from: \url{https://pypi.org/project/hbat/}
\item
  Archived source code: \url{https://doi.org/10.5281/zenodo.XXXXXXX}
\item
  License: \href{https://opensource.org/licenses/MIT}{MIT License}
\end{itemize}

\section{Competing Interests}\label{competing-interests}

No competing interests were disclosed.

\section{Grant Information}\label{grant-information}

This work received no specific grant funding.

\section{Acknowledgements}\label{acknowledgements}

The author thanks the original co-developer Sunil K. Panigrahi and
acknowledges the structural biology community for feedback that guided
the modernization of HBAT.

\section*{References}\label{references}
\addcontentsline{toc}{section}{References}

\phantomsection\label{refs}


\begin{CSLReferences}{0}{0}
\bibitem[\citeproctext]{ref-berman2000protein}
\CSLLeftMargin{{[}1{]} }%
\CSLRightInline{H. M. Berman \emph{et al.}, {``The protein data bank,''}
\emph{Nucleic acids research}, vol. 28, no. 1, pp. 235--242, 2000.}

\bibitem[\citeproctext]{ref-mcdonald1994satisfying}
\CSLLeftMargin{{[}2{]} }%
\CSLRightInline{I. K. McDonald and J. M. Thornton, {``Satisfying
hydrogen bonding potential in proteins,''} \emph{Journal of molecular
biology}, vol. 238, no. 5, pp. 777--793, 1994.}

\bibitem[\citeproctext]{ref-lindauer1996hbexplore}
\CSLLeftMargin{{[}3{]} }%
\CSLRightInline{K. Lindauer, C. Bendic, and J. Sühnel, {``HBexplore--a
new tool for identifying and analysing hydrogen bonding patterns in
biological macromolecules,''} \emph{Computer applications in the
biosciences: CABIOS}, vol. 12, no. 4, pp. 281--289, 1996.}

\bibitem[\citeproctext]{ref-salentin_plip_2015}
\CSLLeftMargin{{[}4{]} }%
\CSLRightInline{S. Salentin, S. Schreiber, V. J. Haupt, M. F. Adasme,
and M. Schroeder, {``{PLIP}: Fully automated protein{\textendash}ligand
interaction profiler,''} \emph{Nucleic Acids Research}, vol. 43, no. W1,
pp. W443--W447, Jul. 2015, doi:
\href{https://doi.org/10.1093/nar/gkv315}{nar/gkv315}.}

\bibitem[\citeproctext]{ref-jubb_arpeggio_2017}
\CSLLeftMargin{{[}5{]} }%
\CSLRightInline{H. C. Jubb, A. P. Higueruelo, B. Ochoa-Montaño, W. R.
Pitt, D. B. Ascher, and T. L. Blundell, {``Arpeggio: {A} {Web} {Server}
for {Calculating} and {Visualising} {Interatomic} {Interactions} in
{Protein} {Structures},''} \emph{Journal of Molecular Biology}, vol.
429, no. 3, pp. 365--371, Feb. 2017, doi:
\href{https://doi.org/10.1016/j.jmb.2016.12.004}{j.jmb.2016.12.004}.}

\bibitem[\citeproctext]{ref-durrant_hbonanza_2011}
\CSLLeftMargin{{[}6{]} }%
\CSLRightInline{J. D. Durrant and J. A. McCammon, {``{HBonanza}: {A}
{Computer} {Algorithm} for {Molecular}-{Dynamics}-{Trajectory}
{Hydrogen}-{Bond} {Analysis},''} \emph{Journal of molecular graphics \&
modelling}, vol. 31, pp. 5--9, Nov. 2011, doi:
\href{https://doi.org/10.1016/j.jmgm.2011.07.008}{j.jmgm.2011.07.008}.}

\bibitem[\citeproctext]{ref-wang_hbcalculator_2024}
\CSLLeftMargin{{[}7{]} }%
\CSLRightInline{Y. Wang and Y. Wang, {``{HBCalculator}: {A} {Tool} for
{Hydrogen} {Bond} {Distribution} {Calculations} in {Molecular}
{Dynamics} {Simulations},''} \emph{Journal of Chemical Information and
Modeling}, vol. 64, no. 6, pp. 1772--1777, Mar. 2024, doi:
\href{https://doi.org/10.1021/acs.jcim.4c00054}{acs.jcim.4c00054}.}

\bibitem[\citeproctext]{ref-siemers_interactive_2021}
\CSLLeftMargin{{[}8{]} }%
\CSLRightInline{M. Siemers and A.-N. Bondar, {``Interactive {Interface}
for {Graph}-{Based} {Analyses} of {Dynamic} {H}-{Bond} {Networks}:
{Application} to {Spike} {Protein} {S},''} \emph{Journal of Chemical
Information and Modeling}, vol. 61, no. 6, pp. 2998--3014, Jun. 2021,
doi:
\href{https://doi.org/10.1021/acs.jcim.1c00306}{acs.jcim.1c00306}.}

\bibitem[\citeproctext]{ref-noauthor_431_nodate}
\CSLLeftMargin{{[}9{]} }%
\CSLRightInline{{``4.3.1. {Hydrogen} {Bond} {Analysis} {\textemdash}
{MDAnalysis}.analysis.hydrogenbonds.hbond\_analysis {\textemdash}
{MDAnalysis} 1.1.0 documentation.''} Accessed: Aug. 08, 2025.
{[}Online{]}. Available:
\url{https://docs.mdanalysis.org/1.1.0/documentation_pages/analysis/hydrogenbonds.html}}

\bibitem[\citeproctext]{ref-noauthor_gmx_nodate}
\CSLLeftMargin{{[}10{]} }%
\CSLRightInline{{``Gmx hbond - {GROMACS} 2025.2 documentation.''}
Accessed: Aug. 08, 2025. {[}Online{]}. Available:
\url{https://manual.gromacs.org/current/onlinehelp/gmx-hbond.html}}

\bibitem[\citeproctext]{ref-noauthor_hbond_2020}
\CSLLeftMargin{{[}11{]} }%
\CSLRightInline{{``Hbond {\textendash} {AMBER}.''} May 2020. Accessed:
Aug. 08, 2025. {[}Online{]}. Available:
\url{https://amberhub.chpc.utah.edu/hbond/}}

\bibitem[\citeproctext]{ref-noauthor_vmd_nodate}
\CSLLeftMargin{{[}12{]} }%
\CSLRightInline{{``{VMD} {HBonds} {Plugin}, {Version} 1.2.''} Accessed:
Aug. 08, 2025. {[}Online{]}. Available:
\url{https://www.ks.uiuc.edu/Research/vmd/plugins/hbonds/}}

\bibitem[\citeproctext]{ref-noauthor_tool_nodate}
\CSLLeftMargin{{[}13{]} }%
\CSLRightInline{{``Tool: {H}-{Bonds}.''} Accessed: Aug. 08, 2025.
{[}Online{]}. Available:
\url{https://www.cgl.ucsf.edu/chimerax/docs/user/tools/hbonds.html}}

\bibitem[\citeproctext]{ref-ferruz_proteintools_2021}
\CSLLeftMargin{{[}14{]} }%
\CSLRightInline{N. Ferruz, S. Schmidt, and B. Höcker, {``{ProteinTools}:
A toolkit to analyze protein structures,''} \emph{Nucleic Acids
Research}, vol. 49, no. W1, pp. W559--W566, Jul. 2021, doi:
\href{https://doi.org/10.1093/nar/gkab375}{nar/gkab375}.}

\bibitem[\citeproctext]{ref-tiwari2007hbat}
\CSLLeftMargin{{[}15{]} }%
\CSLRightInline{A. Tiwari and S. K. Panigrahi, {``HBAT: A complete
package for analysing strong and weak hydrogen bonds in macromolecular
crystal structures,''} \emph{In Silico Biology}, vol. 7, no. 6, pp.
651--661, 2007, doi:
\href{https://doi.org/10.3233/ISI-2007-00337}{ISI-2007-00337}.}

\bibitem[\citeproctext]{ref-noauthor_openmmpdbfixer_2025}
\CSLLeftMargin{{[}16{]} }%
\CSLRightInline{{pdbfixer.''} OpenMM, Aug. 2025. Accessed: Aug.
08, 2025. {[}Online{]}. Available:
\url{https://github.com/openmm/pdbfixer}}

\bibitem[\citeproctext]{ref-eastman_openmm_2013}
\CSLLeftMargin{{[}17{]} }%
\CSLRightInline{P. Eastman \emph{et al.}, {``{OpenMM} 4: {A} {Reusable},
{Extensible}, {Hardware} {Independent} {Library} for {High}
{Performance} {Molecular} {Simulation},''} \emph{Journal of Chemical
Theory and Computation}, vol. 9, no. 1, pp. 461--469, Jan. 2013, doi:
\href{https://doi.org/10.1021/ct300857j}{ct300857j}.}

\bibitem[\citeproctext]{ref-oboyle_pybel_2008}
\CSLLeftMargin{{[}18{]} }%
\CSLRightInline{N. M. O'Boyle, C. Morley, and G. R. Hutchison, {``Pybel:
A {Python} wrapper for the {OpenBabel} cheminformatics toolkit,''}
\emph{Chemistry Central Journal}, vol. 2, no. 1, p. 5, Dec. 2008, doi:
\href{https://doi.org/10.1186/1752-153X-2-5}{1752-153X-2-5}.}

\bibitem[\citeproctext]{ref-mcgaughey_pi-stacking_1998}
\CSLLeftMargin{{[}19{]} }%
\CSLRightInline{G. B. McGaughey, M. Gagné, and A. K. Rappé,
{``\(\pi\)-{Stacking} {Interactions}: {ALIVE} {AND} {WELL} {IN}
{PROTEINS}*,''} \emph{Journal of Biological Chemistry}, vol. 273, no.
25, pp. 15458--15463, Jun. 1998, doi:
\href{https://doi.org/10.1074/jbc.273.25.15458}{jbc.273.25.15458}.}

\bibitem[\citeproctext]{ref-vernon_pi-pi_nodate}
\CSLLeftMargin{{[}20{]} }%
\CSLRightInline{R. M. Vernon \emph{et al.}, {``Pi-{Pi} contacts are an
overlooked protein feature relevant to phase separation,''}
\emph{eLife}, vol. 7, p. e31486, doi:
\href{https://doi.org/10.7554/eLife.31486}{eLife.31486}.}

\bibitem[\citeproctext]{ref-rahim_reciprocal_2017}
\CSLLeftMargin{{[}21{]} }%
\CSLRightInline{A. Rahim, P. Saha, K. K. Jha, N. Sukumar, and B. K.
Sarma, {``Reciprocal carbonyl{\textendash}carbonyl interactions in small
molecules and proteins,''} \emph{Nature Communications}, vol. 8, no. 1,
p. 78, Jul. 2017, doi:
\href{https://doi.org/10.1038/s41467-017-00081-x}{s41467-017-00081-x}.}

\bibitem[\citeproctext]{ref-newberry_n_2017}
\CSLLeftMargin{{[}22{]} }%
\CSLRightInline{R. W. Newberry and R. T. Raines, {``The
n{\textrightarrow}\(\pi\)* {Interaction},''} \emph{Accounts of Chemical
Research}, vol. 50, no. 8, pp. 1838--1846, Aug. 2017, doi:
\href{https://doi.org/10.1021/acs.accounts.7b00121}{acs.accounts.7b00121}.}

\bibitem[\citeproctext]{ref-choudhary_nature_2009}
\CSLLeftMargin{{[}23{]} }%
\CSLRightInline{A. Choudhary, D. Gandla, G. R. Krow, and R. T. Raines,
{``Nature of {Amide} {Carbonyl}-{Carbonyl} {Interactions} in
{Proteins},''} \emph{Journal of the American Chemical Society}, vol.
131, no. 21, pp. 7244--7246, Jun. 2009, doi:
\href{https://doi.org/10.1021/ja901188y}{ja901188y}.}

\bibitem[\citeproctext]{ref-desiraju_weak_2001}
\CSLLeftMargin{{[}24{]} }%
\CSLRightInline{G. Desiraju and T. Steiner, \emph{The {Weak} {Hydrogen}
{Bond}}. Oxford University Press, 2001. doi:
\href{https://doi.org/10.1093/acprof:oso/9780198509707.001.0001}{acprof:oso/9780198509707.001.0001}.}

\bibitem[\citeproctext]{ref-cavallo_halogen_2016}
\CSLLeftMargin{{[}25{]} }%
\CSLRightInline{G. Cavallo \emph{et al.}, {``The {Halogen} {Bond},''}
\emph{Chemical Reviews}, vol. 116, no. 4, pp. 2478--2601, Feb. 2016,
doi:
\href{https://doi.org/10.1021/acs.chemrev.5b00484}{acs.chemrev.5b00484}.}

\bibitem[\citeproctext]{ref-brandl_c-h-interactions_2001}
\CSLLeftMargin{{[}26{]} }%
\CSLRightInline{M. Brandl, M. S. Weiss, A. Jabs, J. Sühnel, and R.
Hilgenfeld, {``C-h?\(\pi\)-interactions in proteins,''} \emph{Journal of
Molecular Biology}, vol. 307, no. 1, pp. 357--377, Mar. 2001, doi:
\href{https://doi.org/10.1006/jmbi.2000.4473}{jmbi.2000.4473}.}

\bibitem[\citeproctext]{ref-hagberg2008networkx}
\CSLLeftMargin{{[}27{]} }%
\CSLRightInline{A. Hagberg, P. Swart, and D. S. Chult, {``Exploring
network structure, dynamics, and function using NetworkX,''} in
\emph{Proceedings of the 7th python in science conference}, Pasadena, CA
USA, 2008, pp. 11--15.}

\bibitem[\citeproctext]{ref-hunter2007matplotlib}
\CSLLeftMargin{{[}28{]} }%
\CSLRightInline{J. D. Hunter, {``Matplotlib: A 2D graphics
environment,''} \emph{Computing in science \& engineering}, vol. 9, no.
3, pp. 90--95, 2007, doi:
\href{https://doi.org/10.1109/MCSE.2007.55}{MCSE.2007.55}.}

\bibitem[\citeproctext]{ref-graphviz2024}
\CSLLeftMargin{{[}29{]} }%
\CSLRightInline{Graphviz Contributors, {``Graphviz - graph visualization
software.''} \url{https://graphviz.org/}, 2024.}

\bibitem[\citeproctext]{ref-tiwari_hbng_2007}
\CSLLeftMargin{{[}30{]} }%
\CSLRightInline{A. Tiwari and V. Tiwari, {``{HBNG}: {Graph} theory based
visualization of hydrogen bond networks in protein structures,''}
\emph{Bioinformation}, vol. 2, no. 1, pp. 28--30, Jul. 2007, doi:
\href{https://doi.org/10.6026/97320630002028}{97320630002028}.}

\bibitem[\citeproctext]{ref-rego_3dmol_2014}
\CSLLeftMargin{{[}31{]} }%
\CSLRightInline{N. Rego and D. Koes, {``3Dmol.js: Molecular
visualization with WebGL,''} \emph{Bioinformatics}, vol. 31, no. 8, pp.
1322--1324, Dec. 2014, doi:
\href{https://doi.org/10.1093/bioinformatics/btu829}{bioinformatics/btu829}.}

\bibitem[\citeproctext]{ref-diaz-cervantes_molecular_2020}
\CSLLeftMargin{{[}32{]} }%
\CSLRightInline{E. Díaz-Cervantes, C. J. Cortés-García, L.
Chacón-García, and A. Suárez-Castro, {``Molecular docking and
pharmacophoric modelling of 1,5-disubstituted tetrazoles as inhibitors
of two proteins present in cancer, the {ABL} and the mutated {T315I}
kinase,''} \emph{In Silico Pharmacology}, vol. 8, no. 1, p. 6, Nov.
2020, doi:
\href{https://doi.org/10.1007/s40203-020-00059-6}{s40203-020-00059-6}.}

\bibitem[\citeproctext]{ref-ravindran_interaction_2015}
\CSLLeftMargin{{[}33{]} }%
\CSLRightInline{R. Ravindran \emph{et al.}, {``Interaction {Studies} of
{Withania} {Somnifera}'s {Key} {Metabolite} {Withaferin} {A} with
{Different} {Receptors} {Assoociated} with {Cardiovascular}
{Disease},''} \emph{Current Computer-Aided Drug Design}, vol. 11, no. 3,
pp. 212--221, 2015, doi:
\href{https://doi.org/10.2174/1573409912666151106115848}{1573409912666151106115848}.}

\bibitem[\citeproctext]{ref-zhou_prediction_2013}
\CSLLeftMargin{{[}34{]} }%
\CSLRightInline{W. Zhou, D. D. Wang, H. Yan, M. Wong, and V. Lee,
{``Prediction of anti-{EGFR} drug resistance base on binding free energy
and hydrogen bond analysis,''} in \emph{2013 {IEEE} {Symposium} on
{Computational} {Intelligence} in {Bioinformatics} and {Computational}
{Biology} ({CIBCB})}, Apr. 2013, pp. 193--197. doi:
\href{https://doi.org/10.1109/CIBCB.2013.6595408}{CIBCB.2013.6595408}.}

\bibitem[\citeproctext]{ref-russo_krauss_thrombinaptamer_2011}
\CSLLeftMargin{{[}35{]} }%
\CSLRightInline{I. Russo Krauss, A. Merlino, C. Giancola, A. Randazzo,
L. Mazzarella, and F. Sica, {``Thrombin--aptamer recognition: A revealed
ambiguity,''} \emph{Nucleic Acids Research}, vol. 39, no. 17, pp.
7858--7867, Sep. 2011, doi:
\href{https://doi.org/10.1093/nar/gkr522}{nar/gkr522}.}

\bibitem[\citeproctext]{ref-wang_simultaneously_2020}
\CSLLeftMargin{{[}36{]} }%
\CSLRightInline{C.-H. Wang, L.-H. Lu, C. Huang, B.-F. He, and R.-B.
Huang, {``Simultaneously {Improved} {Thermostability} and {Hydrolytic}
{Pattern} of {Alpha}-{Amylase} by {Engineering} {Central} {Beta}
{Strands} of {TIM} {Barrel},''} \emph{Applied Biochemistry and
Biotechnology}, vol. 192, no. 1, pp. 57--70, Sep. 2020, doi:
\href{https://doi.org/10.1007/s12010-020-03308-8}{s12010-020-03308-8}.}

\bibitem[\citeproctext]{ref-fournier_vanadium_2014}
\CSLLeftMargin{{[}37{]} }%
\CSLRightInline{J.-B. Fournier \emph{et al.}, {``The {Vanadium}
{Iodoperoxidase} from the {Marine} {Flavobacteriaceae} {Species}
{Zobellia} galactanivorans {Reveals} {Novel} {Molecular} and
{Evolutionary} {Features} of {Halide} {Specificity} in the {Vanadium}
{Haloperoxidase} {Enzyme} {Family},''} \emph{Applied and Environmental
Microbiology}, vol. 80, no. 24, pp. 7561--7573, Dec. 2014, doi:
\href{https://doi.org/10.1128/AEM.02430-14}{AEM.02430-14}.}

\bibitem[\citeproctext]{ref-huang_kernel_2009}
\CSLLeftMargin{{[}38{]} }%
\CSLRightInline{L. Huang, L. Massa, and J. Karle, {``Kernel energy
method applied to vesicular stomatitis virus nucleoprotein,''}
\emph{Proceedings of the National Academy of Sciences}, vol. 106, no. 6,
pp. 1731--1736, Feb. 2009, doi:
\href{https://doi.org/10.1073/pnas.0811959106}{pnas.0811959106}.}

\bibitem[\citeproctext]{ref-jayaprakash_atomic_2025}
\CSLLeftMargin{{[}39{]} }%
\CSLRightInline{N. G. Jayaprakash, D. K. Sarkar, and A. Surolia,
{``Atomic visualization of flipped-back conformations of high mannose
glycans interacting with cargo lectins: {An} {MD} simulation
perspective,''} \emph{Proteins: Structure, Function, and
Bioinformatics}, vol. 93, no. 1, pp. 112--133, 2025, doi:
\href{https://doi.org/10.1002/prot.26556}{prot.26556}.}

\bibitem[\citeproctext]{ref-barzegar_stabilization_2025}
\CSLLeftMargin{{[}40{]} }%
\CSLRightInline{A. Barzegar and N. Tohidifar, {``Stabilization of
{Hoogsteen} {H}-bonds in {G}-quartet sheets by coordinated {K}+ ion for
enhanced efficiency in guanine-rich {DNA} nanomotor,''} \emph{BioImpacts
: BI}, vol. 15, p. 30596, May 2025, doi:
\href{https://doi.org/10.34172/bi.30596}{bi.30596}.}

\bibitem[\citeproctext]{ref-daubner_synanti_2012}
\CSLLeftMargin{{[}41{]} }%
\CSLRightInline{G. M. Daubner, A. Cléry, S. Jayne, J. Stevenin, and F.
H. Allain, {``A syn--anti conformational difference allows {SRSF2} to
recognize guanines and cytosines equally well,''} \emph{The EMBO
Journal}, vol. 31, no. 1, pp. 162--174, Jan. 2012, doi:
\href{https://doi.org/10.1038/emboj.2011.367}{emboj.2011.367}.}

\bibitem[\citeproctext]{ref-kavitha_insights_2024}
\CSLLeftMargin{{[}42{]} }%
\CSLRightInline{K. Kavitha and A. Mohanapriya, {``Insights into the
structure--function relationship of missense mutations in the human
{TOP2A} protein in ovarian cancer,''} \emph{Frontiers in Physics}, vol.
12, Mar. 2024, doi:
\href{https://doi.org/10.3389/fphy.2024.1358406}{fphy.2024.1358406}.}

\bibitem[\citeproctext]{ref-khan_prediction_2017}
\CSLLeftMargin{{[}43{]} }%
\CSLRightInline{I. Khan and I. A. Ansari, {``Prediction of a highly
deleterious mutation {E17K} in {AKT}-1 gene: {An} \emph{in silico}
approach,''} \emph{Biochemistry and Biophysics Reports}, vol. 10, pp.
260--266, Jul. 2017, doi:
\href{https://doi.org/10.1016/j.bbrep.2017.04.013}{j.bbrep.2017.04.013}.}

\bibitem[\citeproctext]{ref-khan_identification_2018}
\CSLLeftMargin{{[}44{]} }%
\CSLRightInline{I. Khan, I. A. Ansari, P. Singh, J. F. P. Dass, and F.
Khan, {``Identification and characterization of functional single
nucleotide polymorphisms ({SNPs}) in {Axin} 1 gene: A molecular dynamics
approach,''} \emph{Cell Biochemistry and Biophysics}, vol. 76, no. 1,
pp. 173--185, Jun. 2018, doi:
\href{https://doi.org/10.1007/s12013-017-0818-1}{s12013-017-0818-1}.}

\bibitem[\citeproctext]{ref-abdulazeez_rs61742690_2019}
\CSLLeftMargin{{[}45{]} }%
\CSLRightInline{S. Abdulazeez, S. Sultana, N. B. Almandil, D. Almohazey,
B. J. Bency, and J. F. Borgio, {``The rs61742690 ({S783N}) single
nucleotide polymorphism is a suitable target for disrupting
{BCL11A}-mediated foetal-to-adult globin switching,''} \emph{PLOS ONE},
vol. 14, no. 2, p. e0212492, Feb. 2019, doi:
\href{https://doi.org/10.1371/journal.pone.0212492}{journal.pone.0212492}.}

\bibitem[\citeproctext]{ref-abdulazeez_-silico_2016}
\CSLLeftMargin{{[}46{]} }%
\CSLRightInline{S. AbdulAzeez and J. F. Borgio, {``In-{Silico}
{Computing} of the {Most} {Deleterious} {nsSNPs} in {HBA1} {Gene},''}
\emph{PLOS ONE}, vol. 11, no. 1, p. e0147702, Jan. 2016, doi:
\href{https://doi.org/10.1371/journal.pone.0147702}{journal.pone.0147702}.}

\end{CSLReferences}
\end{document}